\documentclass[pre,aps,twocolumn,showpacs]{revtex4}

\newcommand{\beq}{\begin{equation}}
\newcommand{\eeq}{\end{equation}}
\newcommand{\bqn}{\begin{eqnarray}}
\newcommand{\eqn}{\end{eqnarray}}
\newcommand{\bqns}{\begin{eqnarray*}}
\newcommand{\eqns}{\end{eqnarray*}}
\newcommand{\bary}{\begin{array}}
\newcommand{\eary}{\end{array}}
\newcommand{\non}{\nonumber}

\usepackage{graphicx}%
\usepackage{amsmath}

\begin{document}
\title{\bf Drifting diffusion on a circle as
continuous limit of a multiurn Ehrenfest model}
\author{Pi-Gang Luan}
\affiliation{Institute of Optical Sciences, National Central
University, Chung-Li 32054, Taiwan, Republic of China }
\author{Yee-Mou Kao}
\affiliation{National Center for High-Performance Computing,
No.21, Nan-ke 3rd.Rd., Hsin-Shi, Tainan County 744, Taiwan,
Republic of China}
\date{\today}

\begin{abstract}
We study the continuous limit of a multibox Erhenfest urn model
proposed before by the authors. The evolution of the resulting continuous system is governed by a differential
equation, which describes a diffusion process on a circle with a nonzero drifting velocity.
The short time behavior of this diffusion process is obtained directly by solving
the equation, while the long time behavior is derived using
the Poisson summation formula. They reproduce the previous results in
the large $M$ (number of boxes) limit. We also discuss the connection between
this diffusion equation and the Schr$\ddot{\rm o}$dinger equation of some quantum mechanical
problems.
\end{abstract}

 \maketitle


In a previous study \cite{KL} we proposed a generalized Ehrenfest urn model \cite{Erf}
with $N$ balls and $M$ urns that arranged periodically along a circle. The evolution of the
system is governed by a directed stochastic operation. Using the standard
matrix diagonalization procedures together with a multi-variable generating
function method, we have solved the problem completely. We found that for a generic
$M>2$ case the average number of balls in a certain urn oscillates several times
before it reaches a stationary value. We also obtained the
Poincar\'{e} cycle \cite{Huang}, i.e., the average time interval required for
the system to return to its initial configuration. The result is simply
given by $M^N$, which indicates that the fundamental assumption of statistical
mechanics holds in this system. Taking $M=2$, our model reproduces all
the results of the original Erhenfest urn model \cite{Erf}.

In this paper, we further study the continuous limit (the large $M$ and $N$ limit) of the proposed
multiurn model.  We show that by defining a density function $\rho$ as the continuous limit of
the fraction $f_i=\langle m_i\rangle/N$, i.e., the average number of balls in the $i$th urn
devided by $N$, the continuous limit of the model exists if we also define the drifting velocity
and diffusion constant appropriately. The evolution of $\rho$ in
spacetime is then governed by a differential equation, which can be solved under proper
initial condition and boundary conditions. The results obtained in this paper are in agreement
with those obtained before by the standard matrix diagonalization method.
Since for even a generic $M$-urn and $N$-ball case the Poincar'e cycle $M^N$ is too huge
to be experienced, the evolution of the system can in practice be treated as unrepeatable, thus
the average quantities considered here become more important than those of microstate details.

We start from the Eq.~(4) of Ref.\cite{KL}: \beq \langle
m_i\rangle_{s}=\left(1-\frac{1}{N}\right)\langle
m_i\rangle_{s-1}+\frac{1}{N}\langle m_{i-1}\rangle_{s-1},\eeq which can be rewritten as
\beq
f_i(s)-f_i(s-1)=-\frac{1}{N}\left[f_i(s-1)-f_{i-1}(s-1)\right],\label{ff}\eeq
where $f_i(s)\equiv\langle m_i\rangle_s/N$,
$N$ is the total number of the balls, and $\langle m_i\rangle_s$ denotes
the number of balls in the $i$th urn after $s$ steps. Adding
$[f_{i+1}(s-1)-f_{i-1}(s-1)]/2N$ to both sides of
Eq.~(\ref{ff}), we get \bqn && \frac{f_i(s)-f_i(s-1)}{\Delta
t}+\frac{\Delta x}{N\Delta
t}\left[\frac{f_{i+1}(s-1)-f_{i-1}(s-1)}{2\Delta x}\right]\non\\
&&=\frac{(\Delta x)^2}{2N\Delta
t}\left[\frac{f_{i+1}(s-1)-2f_i(s-1)+f_{i-1}(s-1)}{(\Delta
x)^2}\right],\label{ff1}\eqn where $\Delta t$ represents the time
interval in one step, and $\Delta x$ stands for the center-center distance between
two neighboring urns. Taking the
continuous limit, we obtain \beq \frac{\partial\rho}{\partial
t}+v\frac{\partial\rho}{\partial x}=D\frac{\partial^2
\rho}{\partial x^2},\label{diffusion1}\eeq where we have used the
substitutions: \beq f_i(s)\rightarrow \rho(x,t),\;\;\;\frac{\Delta
x}{N\Delta t}\rightarrow v,\;\;\;\;\frac{(\Delta x)^2}{2N\Delta
t}\rightarrow D.\label{subst}\eeq
It is clear that Eq.~(\ref{diffusion1}) is a diffusion equation.
Since the model is defined on a circle, we replace $x$ by $\phi$, $v$ by
$\omega$, $\Delta x$ by $\theta$, and the diffusion equation
becomes \beq \frac{\partial\rho}{\partial
t}+\omega\frac{\partial\rho}{\partial {\phi}}=D\frac{\partial^2
\rho}{\partial {\phi}^2}.\label{diffusion1a}\eeq

Before further exploring Eq.(\ref{diffusion1a}), here we give a simple
and general derivation of the diffusion equation.
Note that the conservation of probability implies \beq
\frac{\partial\rho}{\partial t} =-\nabla\cdot{\bf
J},\label{rhoj}\eeq where $\rho({\bf r},t)$ is the probability
density and ${\bf J}({\bf r},t)$ is the probability current
density. Now, the probability current can be written as the sum of
two terms, one for the ``diffusion part", and the other for the ``drifting part" of the
probability carriers (the balls). That is \beq {\bf
J}=-D\nabla\rho+\rho{\bf v},\label{dj}\eeq where $D$ is the
diffusion constant and ${\bf v}$ is the drifting velocity caused by
some pumping force.

Substitute (\ref{dj}) into (\ref{rhoj}), we obtain
\beq
\frac{\partial\rho}{\partial t} =D\nabla^2
\rho-\nabla\cdot(\rho{\bf v}).\eeq
We further assume that ${\nabla\cdot\bf v}=0$ (incompressible fluid;
one special case is that ${\bf v}=$ constant), then we have \beq
\frac{\partial\rho}{\partial t} =D\nabla^2 \rho-{\bf
v}\cdot\nabla\rho,\label{dv} \eeq
which is the desired diffusion equation and has the same form as Eq.~(\ref{diffusion1})
and (\ref{diffusion1a}).

On a straight line, the above equation becomes Eq.~(\ref{diffusion1}), and we adopt
the boundary condition \beq \rho(\infty,t)=\rho(-\infty,t)=0.\eeq
On a circle, Eq.~(\ref{dv})
becomes Eq.~(\ref{diffusion1a}), with boundary condition
\beq \rho(\phi,t)=\rho(\phi+2\pi,t).\eeq

Now we find the solutions $\rho$ for the 1D diffusion
equations on a straight line (\ref{diffusion1}) and on a circle (\ref{diffusion1a}), respectively.
Assuming the initially condition:\beq
\rho(x,0)=\delta(x),\label{ini1} \eeq the solution on
a line can be obtained by Fourier transform method \cite{MW}: \beq \rho(x,t)=\frac{1}{\sqrt{4\pi
Dt}}\exp\left[-\frac{(x-vt)^2}{4Dt}\right].\label{linesol}\eeq
Similarly, for the circle problem, given the initial condition \beq \rho(\phi,0)=\delta(\phi),\label{ini2} \eeq
we obtain \beq
\rho(\phi,t)=\frac{1}{\sqrt{4\pi
Dt}}\sum^{\infty}_{n=-\infty} \exp\left[-\frac{(\phi-\omega
t+2n\pi)^2}{4Dt}\right]. \label{circle}\eeq

In deriving Eq.~(\ref{circle}), we have used the identity \beq
\int^{\infty}_{-\infty}f(x)\,dx=
\sum^{\infty}_{n=-\infty}\int^{2\pi}_0f(x+2n\pi)\,dx
\eeq for a localized function $f(x)$, and we have treated the ``circle
problem" as an ``infinite-folded line problem".

\begin{figure}[hbt]
\includegraphics[width=3.2in]{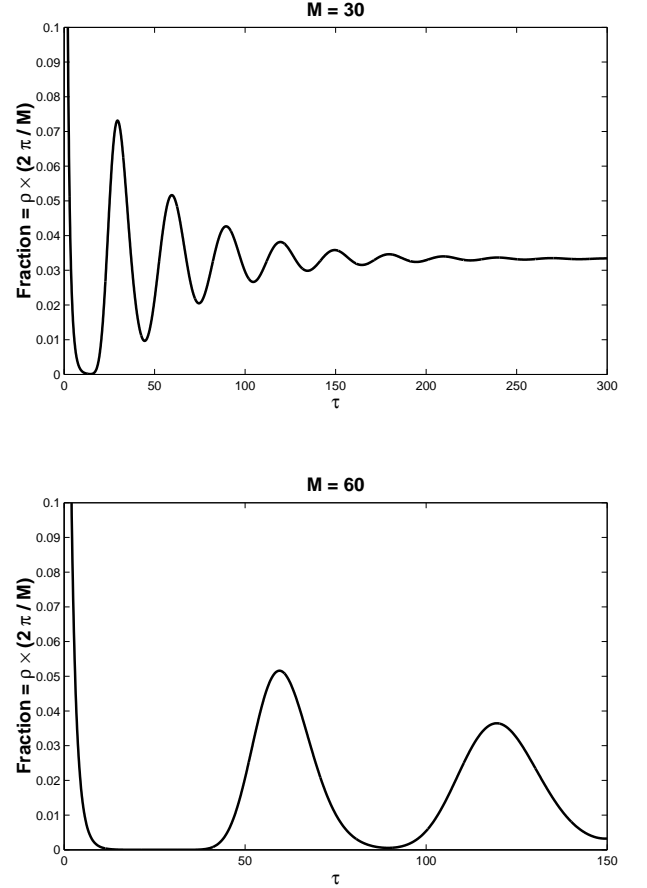}
\caption{\label{figure1}\small Fraction curves calculated from the
exact solution of the diffusion equation.}
\end{figure}

Furthermore, the ``center of mass" now is written
\bqn {\rm COM}
&=&\int^{2\pi}_0\!\!d\phi\,\rho(\phi,t)\, \exp({i\phi}) \non\\
&=& \frac{1}{\sqrt{4\pi
Dt}}\!\sum^{\infty}_{n=-\infty}\!\int^{2\pi}_0\!\!d\phi\, e^{-\frac{(\phi+2n\pi-\omega
t)^2}{4Dt}+i(\phi+2n\pi)}\non\\ &=&\frac{1}{\sqrt{4\pi
Dt}}\int^{\infty}_{-\infty}d\phi\, e^{-\frac{(\phi-\omega
t)^2}{4Dt}+i\phi}\non\\ &=&\frac{e^{i\omega t
}}{\sqrt{4\pi
Dt}}\int^{\infty}_{-\infty}d\phi\, e^{-\frac{\phi^2}{4Dt}+i\phi}\non\\
&=&\exp(-Dt+i\omega t), \label{do}\eqn which is equivalent to Eq.(32) of
Ref. \cite{KL} if we define
\beq
Dt=\frac{\theta^2}{2}\tau=\frac{2\pi^2}{M^2}\tau,\;\;\;\omega
t=\theta\tau=\frac{2\pi}{M}\tau.\label{dtwt}\eeq
Here $\tau$ and $\theta$ are defined as \beq \tau\equiv\frac{t}{N\Delta
t}=\frac{s}{N},\;\;\;\;\theta=\Delta x=\frac{2\pi}{M}.\label{tt}\eeq

Now we compare the results with those in Ref.~\cite{KL}.
Fig.~(\ref{figure1}) shows the results from Eq.~(\ref{circle}), (\ref{dtwt}), and (\ref{tt})
for the cases $M=30$ and $M=60$. As one can see, they indeed reproduce the results of
Ref. \cite{KL} in the large $M$ limit. The parameter $N$ does not appear here because
the motion of each particle is independent in our model.

Note that although for a small $t$ the expression (\ref{circle}) is good enough to
be a fast convergent series, however, when $t$ becomes large, Eq.~(\ref{circle})
converges slowly. In this situation we use a more accurate expression for $\rho$:
\bqn
\rho(\phi,t)&=&\frac{1}{2\pi}\sum^{\infty}_{n=-\infty}
e^{-n^2Dt+in(\phi-\omega t)}\non\\ &=&\frac{1}{2\pi}
+\frac{1}{\pi}\sum^{\infty}_{n=1} e^{-n^2Dt}\cos[n(\phi-\omega
t)],\label{circle1}\eqn which can be derived from (\ref{circle}) using the {\it Poisson summation
formula}: \cite{CH} \beq \sum^{\infty}_{n=-\infty}f(na)=\frac{2\pi}{a}
\sum^{\infty}_{n=-\infty}g\left(\frac{2n\pi}{a}\right).\eeq Here
$f(x)$ is a localized function, and \beq
g(k)=\frac{1}{2\pi}\int^{\infty}_{-\infty}f(x)\,e^{-ikx}\,dx\eeq
is its Fourier transform.

We now consider some solvable generalizations of Eq.~(\ref{diffusion1}) and (\ref{diffusion1a}).
Note that the ratio between $D$ and $\omega$ in our model is fixed:
\beq \frac{D}{\omega}=\frac{\theta}{2}=\frac{\pi}{M}.\label{dw}\eeq
To relax this restriction,
we modify our urn model by assuming that at each time step the picked ball can have probability $p$
to be put into the next urn and probability $q=1-p$ to be put into the previous urn.
Hereafter we call this modified model the $pq$-model. The $pq$-model
is also solvable \cite{KL1} by using methods like those used in Ref. \cite{KL}.
The continuous limit of the $pq$-model can be
derived from the recurrence relation for $f_i$:
\bqn
f_i(s)&=&\left(1-\frac{1}{N}\right)f_i(s-1)\non\\
&+&\frac{p}{N}f_{i-1}(s-1)
+\frac{q}{N}f_{i+1}(s-1),\label{fpq} \eqn where \beq p+q=1,\;\;\;\;{\rm and}\;\;\;\;0\leq p\,,\,q\leq 1.\eeq
Adding
\[\frac{(p-q)}{2N}[f_{i+1}(s-1)-f_{i-1}(s-1)]\] to both sides of
Eq.~(\ref{fpq}), it becomes
\bqn\!\! &&\frac{f_i(s)-f_i(s-1)}{\Delta
t}\non\\&&+\frac{2(p-q)\Delta x}{2N\Delta
t}\left[\frac{f_{i+1}(s-1)-f_{i-1}(s-1)}{2\Delta x }\right]\non\\
&&=\frac{(\Delta x)^2}{2N\Delta
t}\left[\frac{f_{i+1}(s-1)-2f_i(s-1)+f_{i-1}(s-1)}{(\Delta x)^2
}\right].\eqn

Now, define \beq f_i(s)\rightarrow
\rho(x,t),\;\;\;\frac{(p-q)\Delta x}{N\Delta t}\rightarrow
v,\;\;\;\;\frac{(\Delta x)^2}{2N\Delta t}\rightarrow
D,\label{subst1}\eeq then we get a continuous equation of the
form (\ref{diffusion1}), without the restriction (\ref{dw}).
One special case is $p=q=1/2$, which has a zero drifting velocity, and the evolution of the
system is governed by pure diffusion process --- the random walk.

For another generalization we assume that the drifting velocity $v$ varies with time, that is
\beq
\frac{\partial\rho}{\partial t}+v(t)\frac{\partial\rho}{\partial x
}=D\frac{\partial^2 \rho}{\partial x^2}.\eeq Defining $x(t)$ as the
time integral of $v(t)$:
\beq x(t)=\int^t_0 v(t')\,dt',\eeq  and adopting
the initial condition (\ref{ini1}), then
\beq
\rho(x,t)=\frac{1}{\sqrt{4\pi
Dt}}\exp\left[-\frac{\left(x-x(t)\right)^2}{4Dt}\right].\eeq

Similarly, for the diffusion equation on a circle with a
time-dependent $\omega(t)$ and initial condition (\ref{ini2}):
\beq \frac{\partial\rho}{\partial
t}+\omega(t)\frac{\partial\rho}{\partial \phi}=D\frac{\partial^2
\rho}{\partial \phi^2},\label{wt}\eeq and the solution is \beq
\rho(\phi,t)=\frac{1}{\sqrt{4\pi Dt}}\sum^{\infty}_{n=-\infty}
\exp\left[\frac{-(\phi-\phi(t)+2n\pi)^2}{4 Dt}\right].\label{cirsol}\eeq
Here \beq {\phi}(t)=\int^t_0\omega(t')\,dt'.\eeq

The reason for why $v$ and $\omega$ can freely vary with time relies on Eq.(\ref{subst1}).
Recall that in our original multiurn Ehrenfest model or the $pq$-model both the time interval
between two steps and the distance (angle difference) between two urns are undefined.
Thus in deriving the continuous limit of these models we do not have to adopt a constant
$\Delta t$ at each step or a fixed $\Delta x$ ($\Delta \theta$) between two neighboring urns.
If we relax the restriction in Eq.~(\ref{subst1}) and modify them to $\Delta
t_s$ and $\Delta x_i$ ($\Delta \theta_i$), then the continuous limit of these quantities lead
to $v(t)$ or $\omega(t)$.

It is interesting to note that the solutions for the diffusion equation
(\ref{dv}) can be used to find the wave function or Green's function of
some time-dependent quantum mechanical problems \cite{Bauer}.
The main idea is to define a transformation appropriately between the parameters used in the
diffusion equation (\ref{dv}) or (\ref{diffusion1}) and (\ref{diffusion1a}) and those used in the corresponding Schr$\ddot{\rm o}$dinger
equations.
For instance, consider a quantum point particle of charge $q$ and mass $m$ moving under the
influence of a vector potential ${\bf A}(t)$ \cite{Feynman}:
\beq i\hbar\frac{\partial\psi}{\partial t}=\frac{1}{2m}\left(-i\hbar\nabla-\frac{q{\bf A}(t)}{c}
\right)^2\psi,
\label{quant1}\eeq
here we have assumed that ${\bf A}(t)$ is a function of time $t$ only.
Rewriting Eq.~(\ref{quant1}) as
\beq \frac{\partial\psi}{\partial t}=\frac{i\hbar}{2m}\nabla^2\psi+\left(\frac{q{\bf A}}{mc}\right)
\cdot\nabla\psi-\frac{iq^2A^2}{2\hbar mc^2}\psi\eeq
and executing the transformation
\beq \psi=\exp\left[\frac{1}{2i\hbar m}\int^t_0 \left(\frac{qA(t')}{c}\right)^2\,dt'\right]
\tilde{\psi},\label{trans}\eeq
we find
\beq \frac{\partial\tilde{\psi}}{\partial t}=\frac{i\hbar}{2m}\nabla^2\tilde{\psi}+
\left(\frac{q{\bf A}}{mc}\right)\cdot\nabla\tilde{\psi}.\label{quant2}\eeq

Comparing Eq.~(\ref{quant2}) with (\ref{dv}), we find that they can be transformed to each
other by the substitution:
\beq D\leftrightarrow\frac{i\hbar}{2m},\;\;\;\;{\bf v}\leftrightarrow
-\left(\frac{q{\bf A}}{mc}\right),\;\;\;\;
\rho\leftrightarrow \tilde{\psi}.\eeq

To be more specific, consider the case that the particle moving on a
circle of radius $1$. Suppose the circle is lying on the xy-plane and centered at $(x,y)=(0,0)$.
The vector potential can be chosen as ${\bf A}(t)=A(t)\,\hat{e}_{\phi}$ and
is generated by a time-dependent magnetic flux $\Phi(t)$ tube going through
the origin and pointing along the $z$-axis
\beq {\bf A}(t)=A(t)\,\hat{e}_{\phi}
=\frac{\Phi(t)}{2\pi}\,\hat{e}_{\phi}.\eeq

Choosing the initially condition as
\beq \tilde{\psi}(\phi,0)=\psi(\phi,0)=\delta(\phi),\eeq
then
\beq \psi(\phi,t)=\frac{U(t)}{\sqrt{2\pi i\hbar t/m}}\sum^{\infty}_{n=-\infty}
\exp\left[\frac{-(\phi-\phi(t)+2n\pi)^2}{(2 i\hbar t/m)}\right]\label{cirsol}.\eeq
Here
\beq U(t)=\exp\left[\frac{1}{2i\hbar m}\int^t_0 \left(\frac{qA(t')}{c}\right)^2\,dt'\right],\eeq
and \beq \phi(t)=-\frac{q}{mc}\int^t_0 A(t')\,dt'.\eeq

Note that the $\psi$ in Eq.~(\ref{cirsol}) is nothing but the Green's function $G(\phi,\phi_0;t,t_0)$
for the quantum particle with $\phi_0=t_0=0$. If $A(t)=0$, Eq.~(\ref{cirsol}) gives
the well known results \cite{HK}:
\beq G(\phi;t)=\frac{1}{\sqrt{2\pi i\hbar t/m}}\sum^{\infty}_{n=-\infty}\exp
\left[-\frac{(\phi+2n\pi)^2}{(2i\hbar t/m)}\right],\eeq
and
\bqn G(\phi;t)&=&\frac{1}{2\pi}\sum^{\infty}_{n=-\infty}
e^{-n^2(i\hbar/2m)t+in\phi}\non\\
&=&\frac{1}{2\pi}+\frac{1}{\pi}\sum^{\infty}_{n=1}
e^{-n^2(i\hbar/2m)t}\cos n\phi
\eqn
for small and large $t$, respectively.

In conclusion, we have derived the continuous limit of a mutiurn Ehrenfest model,
which is a diffusion equation with a drifting velocity term.
Solving the equation gives us the correct time evolution behavior of the ball distribution.
A transformation was introduced, which changes the solution of the diffusion equation
to the corresponding solution for the problem of a quantum particle moving under the influence
of a time-varying magnetic field.

The supports from NSC, NCTS and NCU are thanked. Discussions
with Dr. D. H. Lin and Dr. C. S. Tang are also acknowledged.


\begin{references}

\bibitem{KL}
Yee-Mou Kao and Pi-Gang Luan, Phys. Rev. E {\bf 67}, 031101, (2003).

\bibitem{Erf}
P. Ehrenfest and T. Ehrenfest, Physik. Z. {\bf 8}, 311 (1907).

\bibitem{Huang}
K. Huang, {\it Statistical Mechanics} (John Wiley \& Sons, Inc,
1987).

\bibitem{MW}
J. Mathews and R. L. Walker, {\it Mathematical methods of physics}, 2nd ed. (Addison-Wesley, Reading, MA,
1971).

\bibitem{CH}
R. Courant and D. Hilbert, {\it Methods of mathematical physics} (New York : Interscience Publishers, Inc,
1953).

\bibitem{KL1}
Yee-Mou Kao and Pi-Gang Luan, unpublished.

\bibitem{Bauer}
J. Bauer, Phys. Rev. A {\bf 65} , 036101 (2002).

\bibitem{Feynman}
R. P. Feynman, R. B. Leighton, and M. L. Sands, {\it The Feynman Lectures on Physics},
Vol. III (Addison-Wesley Publishing Company, 1989)

\bibitem{HK}
H. Kleinert, {\it Path Integrals in Quantum Mechanics, Statistics and Polymer Physics}, 2nd ed.
(World Scientific, Singapore,  1995).

\end{references}
\end{document}